\documentclass[prx, superscriptaddress, reprint]{revtex4-1}
\usepackage[colorlinks,citecolor=blue,linkcolor=blue]{hyperref}
\usepackage{commath}

\newcommand{\be}{\begin{equation}}
	\newcommand{\intf}{\int_{-\infty}^{+\infty}}
	\newcommand{\ee}{\end{equation}}
\newcommand{\bea}{\begin{eqnarray}}
	\newcommand{\eea}{\end{eqnarray}}

\newcommand{\ba}{\begin{array}}
	\newcommand{\ea}{\end{array}}

\newcommand{\bl}{\begin{flalign}}
	\newcommand{\enl}{\end{flalign}}

\newcommand{\pa}{\partial}

\newcommand{\tdse}{time dependent Schr\"{o}dinger equation\ }

\usepackage{siunitx}

\usepackage{graphicx}
\usepackage{amssymb}
\usepackage{epstopdf}
\usepackage{amsmath,dsfont}
\usepackage{bm}
\usepackage{braket}
\usepackage{hyperref, xcolor}
\usepackage[capitalize]{cleveref}

\newcommand{\eq}[1]{Eq.~\eqref{#1}}

\newcommand{\fig}[1]{Fig.~\ref{#1}}
\newcommand{\Fig}[1]{Figure~\ref{#1}}

\newcommand{\tord}{\mathcal{T}}
\newcommand{\tordc}{\mathcal{T}_\gamma}
\newcommand{\intc}{\int_\gamma}

\newcommand*{\rom}[1]{\expandafter\@slowromancap\romannumeral #1@}
\renewcommand{\bf}{\mathbf}
\newcommand{\mc}{\mathcal}

\newcommand{\tr}[1]{\text{Tr}[#1]}

\newcommand{\bs}{\begin{split}}
	\newcommand{\es}{\end{split}}

\renewcommand{\bf}{\mathbf}

\newcommand{\mol}{{(n)}}

\begin{document}
	\title{Toward collective chemistry by strong light-matter coupling}
	\author{Bing Gu}
	\email{gubing@westlake.edu.cn}
	\affiliation{Department of Chemistry \& Department of Physics, School of Science, Westlake University, Hangzhou, Zhejiang 310030, China}
	\affiliation{Institute of Natural Sciences, Westlake Institute for Advanced Study, Hangzhou, Zhejiang 310024, China}

	\begin{abstract}
		Strong light-matter coupling provides a versatile and novel means to manipulate chemical processes.
		Here we develop an exact theoretical framework  to investigate the spectroscopy and dynamics of a molecular ensemble embedded in an optical cavity under the collective strong light-matter coupling regime. This theory is constructed by a pseudoparticle representation of the molecular Hamiltonians, mapping the polaritonic Hamiltonian into a coupled fermion-boson model under particle number constraints. The mapped model is then analyzed using the non-equilibrium Green's function theory with the important self-energy diagrams identified through a power counting. Numerical demonstrations are shown for the driven Tavis-Cummings model, which shows an excellent agreement with  exact results. 
	\end{abstract}

	\maketitle

	\section{Introduction}

	Optical microcavities provide a general platform to manipulate chemical and physical processes of molecules \cite{xiang2020, schwartz2011, garcia-vidal2021, thomas2019, hutchison2012}. The confined cavity photon mode interacts with embedded molecules even without photons. When the light-matter coupling strength surpasses all decay rates of the molecules and the cavity mode, the molecular excitation, either electronic or vibrational, mixes with the cavity excitation to create hybrid light-matter states known as polaritons.   The effective coupling strength between $N$ identical molecules and a cavity mode   scales with the square root of $N$, and inversely with the mode volume $V$, i.e., $\lambda \propto \sqrt{N/V}$. Thus, this leads to two strategies to enhance the coupling strength, by increasing the molecular concentration and by reducing the mode volume.

In the single- and few-molecule $N = 1 \sim 10$ strong coupling regime,  there have been a plethora of theoretical demonstrations that a chemical process can be altered by cavities \cite{mandal2019, gu2020a, Gu2020b, vendrell2018, vendrell2018b, flick2017, flick2018b, galego2019, galego2015, sun2022b, felicetti2020}.
In the optical regime, the optical cavities provide a means to modify the potential energy landscape of molecules, thus can be used to manipulate a photochemical and photophysical process. Both electronic structure theories  such as Hartree-Fock, density functional theory, and coupled cluster, and  quantum dynamics methods, including exact and semiclassical,   haven been extended to incorporate the photonic degrees of freedom, and thus treat polaritonic systems \cite{zhang2019b, flick2018b}.
Specifically, we have shown that the conical intersections in the adiabatic potential energy surfaces can be tuned by coupling to a single cavity mode. This opens the opportunity to control conical interaction-mediated reactions. Examples include internal conversion in pyrazine and singlet fission in pentacene dimer \cite{Gu2020b, Gu2021a, Gu2020d}.  Ground state chemistry can also be altered by coupling a vibrational mode to a resonant cavity mode \cite{li2021, ebbesen2016, hirai2020, li2020a, wang2021d}.

Most of the experimental demonstrations of cavity-altered chemistry uses many $N \gg 1$ molecules to enhance the effective coupling strength to reach the (collective) strong coupling regime \cite{ebbesen2016}.  The intuitive Born-Oppenheimer-like picture that has been developed for single- and few-molecule strong coupling case is not valid for many-molecule case. Besides the two bright polaritonic states, there are a manifold of $N-1$ dark states that can act as  a energy reservoir for the polaritonic states \cite{avramenko2021}. With disorder, the dark states can acquire photonic fractions and are not strictly dark. Understanding the nuclear motion using the highly collective polaritonic surfaces does not always provide an intuitive picture \cite{Gu2020b, perez-sanchez2023}, and it is difficult to include disorder, decoherence, and decay that are  unavoidable in realistic systems.

 Despite substantial efforts,  it is not yet clear under what conditions should we expect the collective strong coupling to alter a chemical process.
 In fact, there are inconsistent experimental results reporting negligible rate changes although polaritonic states are observed \cite{wiesehan2021, imperatore2021}. A crucial step to address this challenge is to have a theoretical and computational framework to describe the polariton dynamics under collective strong coupling. Recen advancements include the molecular dynamics simulation for cavity photochemistry involving more than one thousand molecules \cite{luk2017} , and  a multiconfigurational wavefunction-based method for vibronic models without nonadiabatic couplings in the single-excitation manifold of polariton states, which exploits the permutational symmetry between molecules \cite{perez-sanchez2023}.

Here, we develop a general theoretical framework to understand the collective polariton dynamics under strong light-matter coupling, that can tackle all molecular Hamiltonians including conical intersection models of photochemistry and photophysics, disordered systems, photon leakage, and vibrational Hamiltonians.                                                                                               This theory is exact in the thermodynamic limit $N \rightarrow \infty$. Finite $N$ corrections can be incorporated into the theory by a diagrammatic analysis.
We show that Rabi splitting is a consequence of collective polarization of all molecules.  Using a Tavis-Cummings model under a laser driving, we further show that our theory is in exact agreement with the numerically exact results.

	Our theory starts with a pseudoparticle representation for the molecular Hamiltonian, either a vibronic Hamiltonian for photochemistry and photophysics or a vibrational Hamiltonian for ground state chemistry.
 In the pseudoparticle representation, each molecular eigenstate is mapped to a single-particle orbital. By doing so, the original Hamiltonian can be recast into a coupled fermion-boson model, for which
	the many-body Green's function theory can be applied \cite{Stefanucci2013, Kamenev2011}. The Feynman diagrams serve as a convenient bookkeeping of perturbation series for the self-energy. There are in general infinity diagrams which needs to be taking into account. However, for polaritonic systems, through power counting, the diagrams can be classified by their order of $N^{-1}$.

	In the thermodynamic limit $N \rightarrow \infty$, only diagrams of order $N^0$ is relevant to the polaritonic dynamics. Higher-order diagrams contribute to the finite $N$ corrections. It is shown that the only diagram for the pseudoparticle self-energy that is order $N^0$ is the Hartree diagram, whereas the only diagram important for the photon polarization function is the bubble diagram.

	This paper is organized as follows. The main theory is introduced in \cref{sec:theory}. We first show that the pseudoparticle representation, under a constraint of the particle number, provides equivalent dynamics as the pristine molecular Hamiltonian. This maps the polariton Hamiltonian to a coupled fermion-boson model, which can then be tackled with non-equilibrium Green's function formalism. This is followed by a diagrammatic analysis for the pseudoparticle and photon self-energies. In \cref{sec:discussion}, we discuss how Rabi splitting emerges in the retarded photon Green's function and benchmark the current theory against numerically exact results using a driven Tavis-Cummings model.  \cref{sec:summary} summarises.

	\section{Theory}\label{sec:theory}
	\subsection{The model} \label{sec:gf}
We consider a polaritonic system consisting of $N$ molecules placed inside an optical cavity.	The full Hamiltonian is given by
	\be
	H = \sum_{n=1}^N H_\text{M}^{(n)} + H_\text{C} + H_\text{CM} + H_\text{ext}(t)
\label{eq:h}
\ee
		The molecular Hamiltonian $H_\text{M} = H_\text{BO}(\bf R) + T_\text{n}$, consisting of the electronic Born-Oppenheimer Hamiltonian, that parametrically depends on the nuclear geometry, and the nuclear kinetic energy operator, describes the strongly coupled motion of electrons and nuclei.
	In the  adiabatic representation, the electronic Hamiltonian is diagonalized through standard quantum chemistry methods leading to the adiabatic potential energy surfaces, on which the nuclear wavepackets evolve.
	The cavity Hamiltonian describing a set of cavity photonic modes reads
\be H_\text{C} = \sum_p \omega_p a^\dag_p a_p
\ee
where $p$ labels the cavity modes and $a_p$ ($a_p^\dag$) is the photon annihilation (creation) operator satisfying the bosonic commutation relation $[a_p, a_q] = 0, [a^\dag_p, a^\dag_q] = 0, [a_p, a^\dag_q] = \delta_{pq}$. Here we have neglected the zero-point energy.

The cavity-molecule interaction in the electric-dipole approximation is given by \cite{Andrews2018, Stokes2022, Cohen-Tannoudji1989}
\be
H_\text{CM} = \sum_n - \varepsilon_0^{-1}\bm \mu_n \cdot \hat{\bf D}(\bf r_n)
\ee
where $\bm \mu_n$ is the dipole operator of the $n$th molecule, $\hat{\bf D}(\bf r) = \sum_p i \sqrt{\frac{\varepsilon_0 \omega_p}{2V} } \hat{\bf e} u_p(\bf r){a_p } + \text{H.c.}$ is the displacement field operator and H.c. stands for Hermitian conjugate, and $\epsilon_0$ is the vacuum permittivity.  Here $u_p(\bf r)$ is the mode function satisfying the Maxwell's equations subject to the  boundary conditions set by the mirrors \cite{glauber1991}.  For Fabry-Perot cavities, $ \bf u_p(z) =  \sin(k_p z)$.
Under the long-wavelength approximation, the electric field variation across the molecular sample is neglected. We do not invoke the commonly used rotating wave approximation so that the influence of counterrotating terms are included.


	\subsection{Pseudoparticle representation}
\begin{figure}[htbp]
	\includegraphics[width=0.48\textwidth]{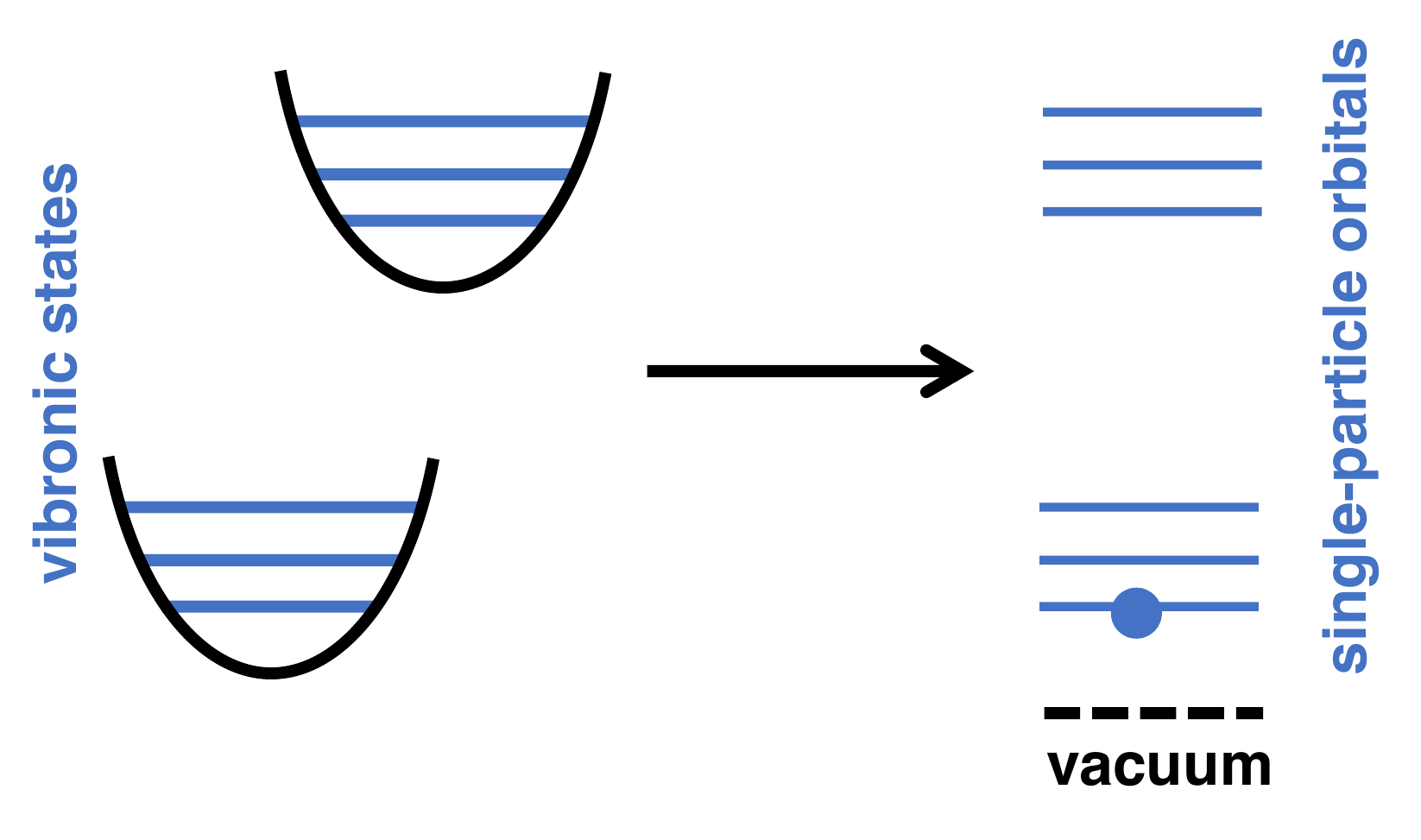}
	\caption{Schematic of the pseudoparticle representation of molecular states. Transitions between vibronic states is mapped to the spinless pseudoparticle hopping between single-particle orbitals. A vacuum state can be defined by the annihilation operators. The constraint of particle number ensures that there is always only one particle occupying the orbitals coming from a single molecule. Thus, the exchange statistics is irrelevant in the pseudoparticle representation. }
	\label{fig:map}
\end{figure}

	Our first step to develop a general many-body theory of polaritonic dynamics is the  pseudoparticle representation of the molecular.
	Denote the molecular eigenstates of $H_\text{M}$ as $\ket{\Phi_\alpha}$,
	i.e., $ H_\text{M}^{(n)}   \ket{\Phi_\alpha^\mol}  = E_\alpha  \ket{\Phi_\alpha^\mol}
	$,
the pseudoparticle representation is achieved by  \cite{fresard2012}
	\be
	X_{\alpha \beta}^\mol \equiv \ket{\Phi_\alpha^\mol}\bra{\Phi_\beta^\mol} \rightarrow c_{n\alpha}^\dag c_{n\beta}
	\ee
		under the constraint of pseudoparticle number
	\be
	\sum_\alpha c_{n\alpha}^\dag c_{n\alpha} = 1,
	\label{eq:constraint}
	\ee
		for any $n \in [1, \cdots, N]$, where $\alpha$ runs over all vibronic states for each molecule.
		The molecular Hamiltonian can be electronic, vibrational or vibronic, Hamiltonian depending on the process of interest.
		For example, it can be a conical intersection model with strong vibronic coupling for photoisomerization or a double well potential for proton tunneling.
		An intuitive way to understand this representation is that each molecular state maps to a single-particle orbital where the pseudoparticle can occupy, schematically shown in \cref{fig:map}. Transitions between molecular states maps to the pseudoparticle hopping between the single-particle orbitals.
	The constraint,  arising from the completeness relation $\sum_\alpha X_{\alpha \alpha} = \bf I$,  ensures that there is always one pseudoparticle in all the single-particle orbitals belonging to the same molecule.
		Due to this constraint of pseudoparticle number, whether the pseudoparticle is fermion or boson becomes irrelevant. One can freely choose the exchange statistics of the pseudoparticle for convenience.
		%

	The pseudoparticle dynamics described by the mapped Hamiltonian is equivalent to the vibronic dynamics under the molecular Hamiltonian, see \cref{fig:map}.
	To show this, it suffices to show that  the commutation relation

	\be
	[X_{\alpha \beta}, X_{\gamma \delta}] = \delta_{\beta \gamma} X_{\alpha \delta} - X_{\gamma \beta}\delta_{\delta \alpha}
	\label{eq:112}
	\ee
	is preserved after the mapping.
	For the mapped Hamiltonian,
	\be
	[c_\alpha^\dag c_\beta, c_\gamma^\dag c_\delta] = c_\alpha^\dag c_\beta c_\gamma^\dag c_\delta -    c_\gamma^\dag c_\delta c_\alpha^\dag c_\beta
	\label{eq:111}
	\ee
	The pseudoparticle can be either bosons or fermions satisfying the canonical commutation or anticommutation relations
	\be
	[c_\alpha, c_\beta^\dag]_\pm =  \delta_{\alpha \beta}
	\label{eq:ccr}
	\ee
	Using \eq{eq:ccr} in \eq{eq:111} yields
	\be
	[c_\alpha^\dag c_\beta, c_\gamma^\dag c_\delta] = c_\alpha^\dag \delta_{\beta \gamma} c_\delta -    c_\gamma^\dag \delta_{\delta \alpha} c_\beta,
	\ee
	consistent with \eq{eq:112}. Therefore, the exchange statistics of the pseudoparticles are irrelevant in the mapping. Here we have chosen pseudoparticles spinless, but one can envision cases where spinful pseudoparticles can be useful. For example, when the eigenstates are degenerate due to e.g. time-reversal symmetry.

	With the pseudoparticle mapping,  the Hamiltonian for all molecules can be expressed as
	\be
	H_\text{M} = \sum_n \sum_{\alpha} E_\alpha c_{n \alpha}^\dag c_{n \alpha}
	\ee
The cavity-molecule coupling is recast as (supposing the coupling strength is homogeneous)
	\be
	H_\text{CM} 
	= \sum_{n=1}^N \sum_{\alpha,\beta} \frac{\lambda_{\alpha \beta}^p }{\sqrt{N}}    c_{n \alpha}^\dag c_{n\beta} \phi_p
	\ee
	where $\phi_p = \frac{1}{\sqrt{2}} \del{a_p + a_p^\dag}$ is the displacement field operator with conjugate momentum $\pi_p$.
		The factor $1/\sqrt{N}$ is to ensure that the Rabi spliting is finite (.i.e. $\mc{O}(N^0)$) in the thermodynamic limit $N \rightarrow \infty$.
	The full mapped Hamiltonian now reads
	\be
	\begin{split}
	H &= \sum_n  \sum_{\alpha} h_{\alpha \beta}(t) c_{n\alpha}^\dag c_{n\beta} + \sum_{p} \frac{\omega_p}{2} \del{\phi_p^2 + \pi_p^2}  \\
	&+ F_p(t)\phi_p(t) + \sum_n \sum_p \sum_{\alpha, \beta} \frac{\lambda_{\alpha \beta}^p }{\sqrt{N}} c_{n\alpha}^\dag c_{n\beta}  \phi_p
	\end{split}
	\ee
where we have combined the bare molecular Hamiltonian and external lasers driving the molecules into the first term.

	\subsection{Non-equilibrium Green's function}
			Here we choose the pseudoparticle to be fermion such that the mapped Hamiltonian describes coupled fermions and bosons. A general theoretical framework to treat such systems is the non-equilibrium many-body Green's function formalism \cite{Stefanucci2013}.   The contour-ordered non-equilibrium GFs, defined in a Keldysh contour, consists of 
			several components. The retarded GF encodes the response function of the system, whereas the lesser component contains the information of particle distribution.

The pseudoparticle GF is defined as
	\be
	G_{\alpha \beta}^{nm}(\tau, \tau')  = -i \Braket{ \tordc c_{n\alpha}(\tau) c_{m\beta}^{\dag}(\tau')}
	\ee
	where $\gamma$ is the Keldysh contour consisting of two branches, the upper branch going from $t = -\infty+i\eta$ to $t = \infty + i\eta$, and the lower branch from $t = +\infty-i\eta$ to $-\infty-i\eta$,  $\tord_\gamma$ is the contour-ordering operator, $\eta > 0$ is an infinitesimal number, $\tau$ is a contour variable.

	An important relation that arises from the constraint \eq{eq:constraint}  is that the inter-molecule Green's function all vanish, i.e., for $n \ne m$,
	\be
	G^{nm}_{\mu \nu}(t,t') = -i \Braket{ \tordc c_{n\mu}(t) c_{m\nu}^\dag(t')} = 0.
	\ee
	The reason is that the pseudoparticle occupying one orbital mapped from one molecule cannot propagate to an orbital belonging to another molecule due to the constraint.
	%
	 The vanishing of inter-molecular GF simplifies drastically the analysis of the model. Supposing the molecules are identical,  the molecular index can be suppressed, $G^{nn} = G$.

	The photon GF is defined similarly

	\be D_{pq}(\tau,\tau') = -i \Braket{ \tordc \Delta \phi_p(\tau) \Delta \phi_q(\tau')}
	\ee
	where $\Delta \phi_p = \phi_p(t) - \braket{\phi_p(t)}$.

	The Heisenberg equation of motion for the displacement field operator is
	\be
	\omega_p^{-1}\del{-\pa_t^2 - \omega_p^2} {\phi_p(t)} =    F_p(t) + \sum_{\alpha, \beta} \lambda_{\alpha \beta}^p c_\alpha^\dag(t) c_\beta(t)
	\label{eq:eom_phi}
	\ee
	Thus,
	\be
	\phi_p(\tau) = \intc \dif \tau'  D_0^p(\tau, \tau') \del{ F_p(\tau) - i \tr{\bm \lambda_p \bf G(\tau,\tau^+)}}
	\ee
	where $ D_0^{p, -1}(\tau, \tau') = \omega_p^{-1}\del{-\pa_\tau^2 - \omega_p^2} \delta(\tau, \tau')$, $\bm \lambda_p$ is  a matrix with elements $\lambda_{ij}^p$.
To describe the polariton dynamics amounts to solving the pseudoparticle and photonic GFs together.
	%
		Hereafter, we consider a single cavity mode such that the photon mode index can be suppressed. It is  straightforward to extend the following analyses to the multimode case.

	Within the many-body GF theory, the Dyson equation in the Keldysh contour are given by
	\be
	\intc \dif 2 \bf G_0^{-1}(1', 2)\bf G(2, 1)
  = \delta(1',1) +
	\intc \dif 2 \bf \Sigma(1', 2)  \bf G(2, 1)
	\ee
	where $\bf G_0$ is the bare pseudoparticle GF with $\bf G_0^{-1}(\tau, \tau') = \delta(\tau, \tau') \del{i\pa_\tau - \bf h(\tau) - v_\text{ext}(\tau)}$. The external potential $v_\text{ext}^{ij}\del{\tau} = \lambda_{ij} \intc D_0(\tau, \tau')F(\tau')$ comes from an external force driving the cavity mode.
	The Dyson equation for the photon propagator reads
	\be
	\intc \dif 2 D_0^{-1}(1, 2)  D(2,1') = \delta(1, 1') +  \intc \dif 2 \Pi(1, 2)  D(2,1') \ee
	where we abbreviate $i \equiv \tau_i$,
	 %
	$\bf \Sigma$ is the pseudoparticle self-energy and $\Pi$ is the polarization function.  Since we are only considering a single cavity mode, the photon GF is a scalar.

	With the help of Langreth rules \cite{Stefanucci2013}, the Dyson equations for the pseudoparticle GFs in real time can be written as
\be
\begin{split}
\del{i\pa_t - \bf h(t) - v_\text{ext}(t) } \bf G^R(t, t') &= \delta(t,t') \bf I \\
&+ \intf \dif s \bf \Sigma^R(t,s) \bf G^R(s, t')
\label{eq:main1}
\end{split}
\ee
and for the lesser GF
\be
\begin{split}
\del{i\pa_t - \bf h(t) - v_\text{ext}(t)} \bf G^<(t, t') &=  \intf \dif s \Sigma^R(t,s) \bf G^<(s, t') \\
& + \intf \dif s \Sigma^<(t,s) \bf G^A(s, t')
\end{split}
\label{eq:main2}
\ee

Similarly,  the equations of motion for the $D^\text{R}$ and $D^<$ reads
\be
\begin{split}
\intf \dif s D_0^{-1}(t, s) D^\text{R}(s, t') &= \delta(t-t') \\
&+
\intf \dif s \Pi^\text{R}(t,s) D^\text{R}(s, t')
\label{eq:main3}
\end{split}
\ee
and
\be
\begin{split}
\intf \dif s D_0^{-1}(t,s) D^<(s,t') &= 
\intf \dif s \Pi^R(t,s) D^<(s, t') \\
&+ \intf \dif s \Pi^<(t,s) D^A(s, t')
\label{eq:main4}.
\end{split}
\ee

 \cref{eq:main1,eq:main2,eq:main3,eq:main4} consist of a coupled set of equations. The self-energy depends on the photon GFs and the polarization function depends on the pseudoparticle  GFs.
	Once the pseudoparticle GFs are known,
	the local observables including chemical reactions can all be obtained from the lesser GF because the intramolecular density matrix $\rho(t) = - i G^<(t, t)$. 	The problem now is  to identify the  Feynman diagrams for the pseudoparticle self-energy  and polarization function, that are important for the collective polariton dynamics.

	\subsection{Power counting} 



To identify relevant diagrams for the polaritonic dynamics, we classify the diagrams by their order with respect to the number of molecules $N^{-1}$. Each vertex contributes a factor of $N^{-1/2}$,  whereas a sum of all molecules contributes a factor of $N$.


\begin{figure}[htbp]
	\includegraphics[width=0.48\textwidth]{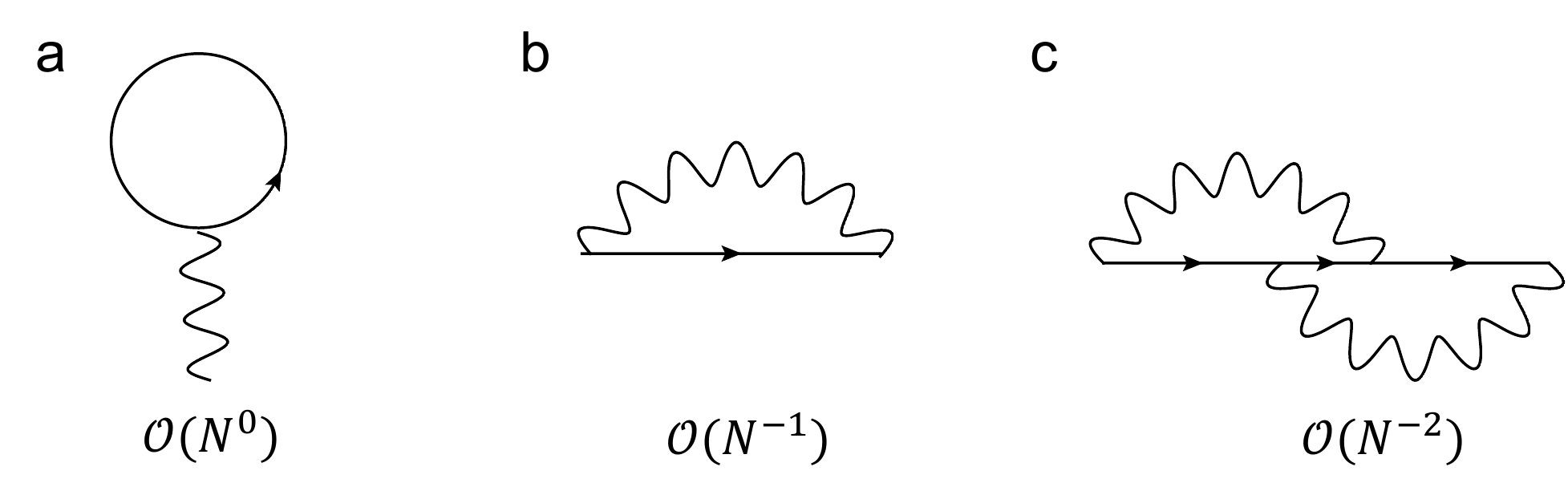}
	\caption{Feynman diagrams for the polarization function. Their order with respect to $N$ is indicated below.}
	\label{fig:diagram}
\end{figure}

\begin{figure}[htbp]
	\includegraphics[width=0.45\textwidth]{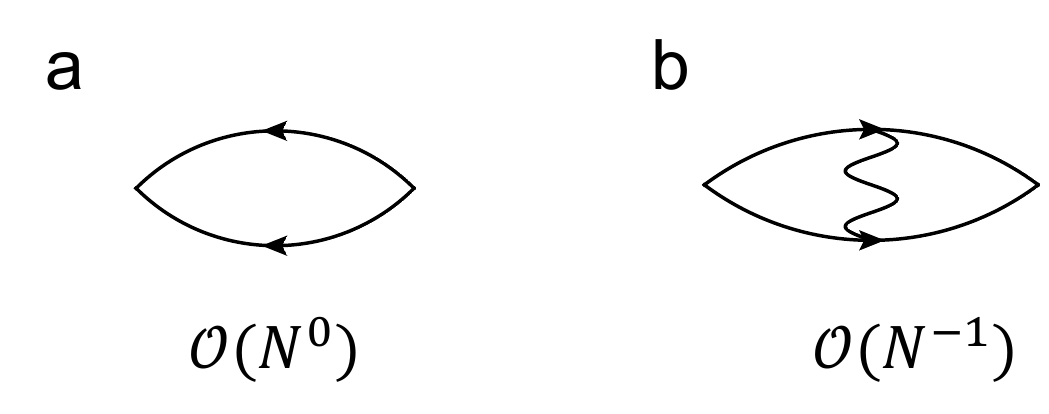}
	\caption{Feynman diagrams for the pseudoparticle self-energy. Their order with respect to $N$ is indicated below.}
	\label{fig:pol}
\end{figure}

The possible Feynman diagrams for the self-energies are depicted in \fig{fig:diagram}, with the order of each diagram indicated below.
The important observation from the power counting analysis  is  that there is only a single diagram of order $N^0$  that can possibly contribute to each self-energy in the thermodynamic limit. For the pseudoparticle, this is the Hartree diagram depicted in  \fig{fig:diagram}a. For the photon self-energy, it is the bubble diagram in \fig{fig:pol}a .
Since we have employed the mapping Hamiltonian, our electronics represent the molecular eigenstates. So the meaning of this diagram differs from the purely electronic Hartree diagram coming from the Coulomb interaction.

Using the Feynman rules, the  polarization function associated with the bubble diagram reads
\be
\Pi(1,2) = -i \tr{\bf G(1,2) \bm \lambda \bf G(2,1) \bm \lambda}
\ee
The Hartree self-energy is given by \cite{sakkinen2015, sakkinen2015a}
\be
\Sigma_\text{H}(\tau, \tau') = \delta(\tau, \tau') v_\text{H}(\tau)
\ee
with the time-local Hartree potential given by
\be v_\text{H}(\tau) =  -i \intc \dif \tau_1  D_0(\tau, \tau_1) \bm \lambda \tr{\bf G(\tau_1,\tau^+_1) \bm \lambda}
\ee
where $\tau_1^+$ represents a time infinitesimally later than $\tau_1$ in the contour.
Here the bare photon propagator is used instead of the dressed one to avoid double counting \cite{ sakkinen2015, Stefanucci2013, Hedin1965}.

With the Hartree self-energy, the equation of motion for the reduced density matrix
\be
i \pa_t \rho(t) =    [\bf h(t) + v_\text{ext}(t) + v_\text{H}(t), \rho(t)]
\label{eq:main}
\ee
with the Hartree potential given by
\be
v_\text{H}(t)  =  \int_0^t \dif t'  D_0(t, t') \bm \lambda \tr{\bf \rho(t') \bm \lambda}
\ee
At order $N^0$, we essentially end up with a mean-field-like theory.

A caveat is on the initial condition, which cannot be the vacuum state due to the constraint. It has to be chosen such that the (modified)  Wick's theorem can be invoked. If the molecule is initially in a single state described by $c$, i.e., $c^\dag\ket{\Omega}$ in the mapped space, a simple trick is to define the pseudohole operator $h = c^\dag$ such that the initial state is annihilated by both $d$ and $h$. Note that the choice of an initial state at initial time does not prevent us from treating non-equilibrium processes  because we have  included external driving in the polariton Hamiltonian.

The retarded  GF is given by
\be
\bf G^{\text{R}}(t, t') = -i \theta(t-t') U(t,t')
\ee
with the propagator
$
U(t,t') = \tord e^{-i \int_{t'}^t \dif s  {\bf h_\text{eff}(s)} }
$.
where we have defined an effective Hamiltonian $\bf h_\text{eff}(t) = \bf h(t) + v_\text{ext}(t) + v_\text{H}[\rho(t)](t)$.
We have indicated the dependence of the Hartree potential on the molecular density matrix. 
The formal solution of the lesser GF can be given by 
\be
G^<_{\alpha \beta}(t, t') = U(t, t_0)\rho_{\alpha \beta}(t_0) U^\dag(t', t_0) .
\ee
where the density matrix $\rho_{\alpha \beta}(t_0) = \braket{c_\beta^\dag c_{\alpha} }$ characterizes the initial particle distribution.

As the molecular density matrix is determined by the lesser GF, i.e., $\rho(t) = -i \bf G^<(t,t)$, whether the local dynamics of each molecule differs from the bare molecular dynamics boils down to whether the Hartree potential contributes to the polaritonic dynamics.

%

\section{Discussion} \label{sec:discussion}
\subsection{Collective effects in the molecular dynamics}

To understand whether there is collective effects on the single-molecule level, it suffices to examine weather the Hartree potential contributes to the polariton dynamics at the thermodynamic limit. \eq{eq:main} implies that inside a cavity, each molecule feels a potential due to the polarization of all molecules  through the bare photon propagator.  Alternatively, realizing that \be v_\text{ext}(t) + v_\text{H}(t) = \bm \lambda \phi(t), \ee
each molecule feels the local cavity field dressed by all molecules.
Whether the polaritonic dynamics will differ from the bare molecular dynamics relies on whether a process generates polarization during the course of dynamics. For example, during the passage through a conical intersection, electronic coherence may emerge  that can create a polarization \cite{kowalewski2016}. 




To illustrate the utility of the pseduparticle non-equilibirum Green's function method (PP-NEGF), we study the Tavis-Cummings (TC) model where exact numerical results can be obtained for finite $N$.   The TC model can be obtained by only including two states in the molecular Hamiltonian followed by invoking the  rotating-wave approximation (RWA),
\be
H_\text{TC} =  \omega_\text{c} a^\dag a + \sum_n \omega_0 \sigma_n^\dag \sigma_n + \frac{\lambda}{\sqrt{N}} \del{\sigma_n a^\dag + \sigma^\dag_n a}
\ee
Since we only have two states, we abbreviate $\lambda_{eg} = \lambda$.

\begin{figure}[htbp]
	\includegraphics[width=0.5\textwidth]{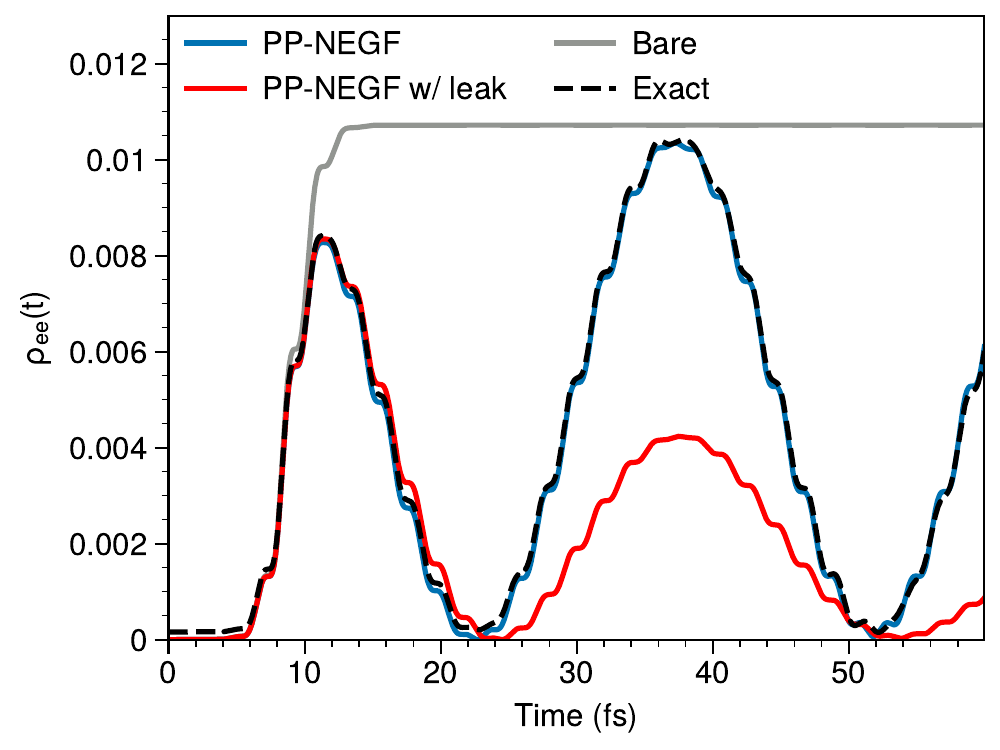}
	\caption{The driven polaritonic dynamics (blue) using the PP-NEGF method  almost overlaps with the numerically exact results (black dashed). The polaritonic dynamics clearly differs from the bare dynamics (grey). For the numerically exact results, $N = 8$, the Fock space of the cavity mode is truncated to 4. }
	\label{fig:pp}
\end{figure}

We contrast the laser-driven polaritonic dynamics computed with the PP-NEGF method to the exact numerical result obtained by directly integrating the \tdse for the TC model. A Gaussian pulse is employed to drive the molecules,
\be
\mc{E}(t) = \mc{E}_0 \cos(\omega t) e^{- (t- T)^2/2\tau^2}
\ee
where $\mc{E}_0$ is the electric field strength,  $T$ is the central time of the pulse, $\tau$ the duration, $\omega$ the carrier frequency.  As shown in \fig{fig:pp}, the driven polaritonic dynamics using the PP-NEGF method  almost overlaps with the numerically exact results. The polaritonic dynamics clearly differs from the bare dynamics.

The cavity decay (i.e., photon leakage to the extracavity modes)  can be incorporated into the theory by introducing an additional polarization function into the photon GF $\Pi^\text{R}(t,t') = -2i \kappa \delta(t,t')$, where $\kappa$ is the decay rate, representing a Markovian decay due to the coupling between the discrete intracavity modes and the continuum of extracavity modes. An illustrative computation is shown in \fig{fig:pp}, where, as expected, a damping of the excited state population is observed.

\subsection{Collective effects in the photon dynamics}
It is known that there is collective effects in the cavity dynamics because the Rabi splitting observed in experiments is of order $N^{1/2}$. We now demonstrate  how this cooperativity can be understood in the current theoretical framework.



Due to the neglect of counter-rotating terms in the TC model, this model has a symmetry that preserves the number of excitations, $[N_\text{ex}, H_\text{TC}] = 0$ where the excitation number operator $N_\text{ex} = a^\dag a + \sum_n \sigma^\dag_n \sigma_n$.
Consequently, it is straightforward to obtain the single-polariton eigenstates of the Tavis-Cummings model as a superposition of the cavity excitation and collective molecular exaction. For simplicity, consider the resonant case $\omega_0 = \omega_\text{c}$, the polariton states are given by $\ket{\text{P}_\pm} = \frac{1}{\sqrt{2}} \del{ \ket{1} + \ket{E}}$ where $\ket{E} = \frac{1}{\sqrt{N}} \sum_n \sigma_n^\dag \ket{0}$ is the bright exciton state. The Rabi splitting $\Omega_\text{R} = 2\lambda$.

It is instructive to see how this exact result can be obtained in the mapping Hamiltonian approach. The ground state (excited state) is mapped to a single-particle orbital described by $c, c^\dag$ ($d, d^\dag$) with energies $0$ ($\omega_0$). The lowering operator becomes $\sigma_n \rightarrow c_n d^\dag_n$.
The corresponding mapping Hamiltonian under the rotating-wave approximation is given by
\be
 H_\text{TC} =   \omega_\text{c} a^\dag a +   + \sum_n \omega_0 d_n^\dag d_n+  \sum_n  \frac{\lambda}{\sqrt{N}} \del{d_n^\dag c_n a + \text{H.c.}}
\ee
The ground state  is simply the product state of all molecular ground states and the vacuum state of the cavity mode, $\ket{G} = \ket{g_1 g_2 \cdots g_N} \otimes \ket{0}$. In the mapped space, $\ket{G} = \prod_n c_n^\dag \ket{{\Omega}} \otimes \ket{0}$, where $\ket{{\Omega}}$ refers to the pseudoparticle vacuum state.

We define pseudo-hole operators as $c_n = h_n^\dag$, such that the Wick's theorem can be used, i.e., the pseudoparticle vacuum is annihilated by both $d_n$ and $h_n$ . The cavity-molecule interaction becomes  $\sum_n  \lambda/\sqrt{N} \del{d_n^\dag h^\dag_n a + \text{H.c.}}$. This leads to the picture of a cavity photon creating a particle-hole pair.

We calculate the cavity transmission spectrum, typically employed to measure the Rabi splitting. A weak external probe impinges the cavity, the transmitted light from the other side of the cavity is recorded. Without matter, the transmission determines the cavity modes.
The interaction between the external laser and light can be modeled by $\mc{E}^*(t) a + \mc{E}(t)a^\dag$. The transmission spectrum is given by the retarded photon GF. Under the RWA, it is more convenient to use the creation and annihilation operators to define the photon GF, i.e.,
\be
F(t,t') = -i \braket{\tord a(t) a^\dag(t')} .
\ee
Since we are considering the linear response of an equilibrium system, we can use equilibrium GF theory with time-translational invariance, i.e.,   $D(t, t') = D(t-t')$.

The polarization function corresponding to the bubble diagram in given by
\be \Pi(t,t') = i \lambda^2 g_d(t,t') g_h(t,t')
\ee
where $g_d(t,t') = - i \braket{\tord d(t) d^\dag(t')}$ and $g_h$ are respectively the bare GF for the particle and hole,

\be
g_d(\omega) =  \frac{1 }{\omega + i\eta - \omega_0} 
\ee
with $\eta > 0$  an infinitesimal number.


The cavity transition spectrum is given by the retarded photon GF. In the frequency domain, the Dyson equation is given by
\be
F^\text{R}(\omega) = \frac{1}{ \del{ F_0^{\text{R}}(\omega)}^{-1} - \Pi^\text{R}(\omega)}
\label{eq:120}
\ee

Using $D_0^\text{R}(\omega) = \frac{1}{\omega - \omega_\text{c} + i \eta}$ in \cref{eq:120}, the pole of the photon GF can be obtained by solving $\del{\omega - \omega_0}\del{\omega - \omega_\text{c}} - \lambda^2 = 0$, which is precisely the secular equation of $H_\text{TC}$ within the single-excitation space $N_\text{ex} = 1$, and thus yields the same eigenvalues.

Thus, we have recovered the exact results for the single-polariton excitations, despite that we have used the bare Green's function for the pseudoparticles.
Generalizing to $M$ molecular states described by $\set{c = h^\dag, d_1, d_2,\cdots, d_{M-1}}$ and considering that the molecule is in the ground state initially $c^\dag \ket{0}$, the polarization function becomes
\be
\Pi(t,t') = i \sum_{i=1}^{M-1} {\lambda_{i0}  g^d_{ii}(t,t') \lambda_{0i}  g^h(t,t')}
\ee

\subsection{Disorder}

Disorder is inevitable for molecular polaritons as each molecule can have a different local environment. Energetic disorder has been shown to be important for their transport and localization properties \cite{engelhardt2022, engelhardt2023}. To include static disorder in the transition frequency (i.e., the transition frequency of each molecule $\omega_n$ becomes a random variable with probability distribution $\varrho(\omega)$), the disorder-averaged GF can be introduced
\be
\overline{F}(\omega) = \frac{1}{F_0^{-1}(\omega) - \overline{\Pi}(\omega)}
\ee
where $\Pi(\omega) = \int \dif \omega' \varrho(\omega') \frac{\lambda^2}{\omega - \omega' + i \gamma}$.

With a Gaussian distribution  $\varrho(\omega) = \mc{N}(\omega_0, \sigma^2)$,
the polarization function becomes

\be
\Pi(\omega) =
\frac{\sqrt{\pi/2}}{ \sigma} e^{-\frac{ (\omega - \omega_0)^2}{2 \sigma ^2}} \left( -i +  \text{erfi} \left( (\omega - \omega_0) / \sqrt{2} \sigma \right) \right)
\label{eq:pgf}
\ee
where $\text{erfi}(x) = -i  \text{erf}(ix)$ is the imaginary error function.

%

\begin{figure}[ht]
	\centering
	\includegraphics[width=0.48\textwidth]{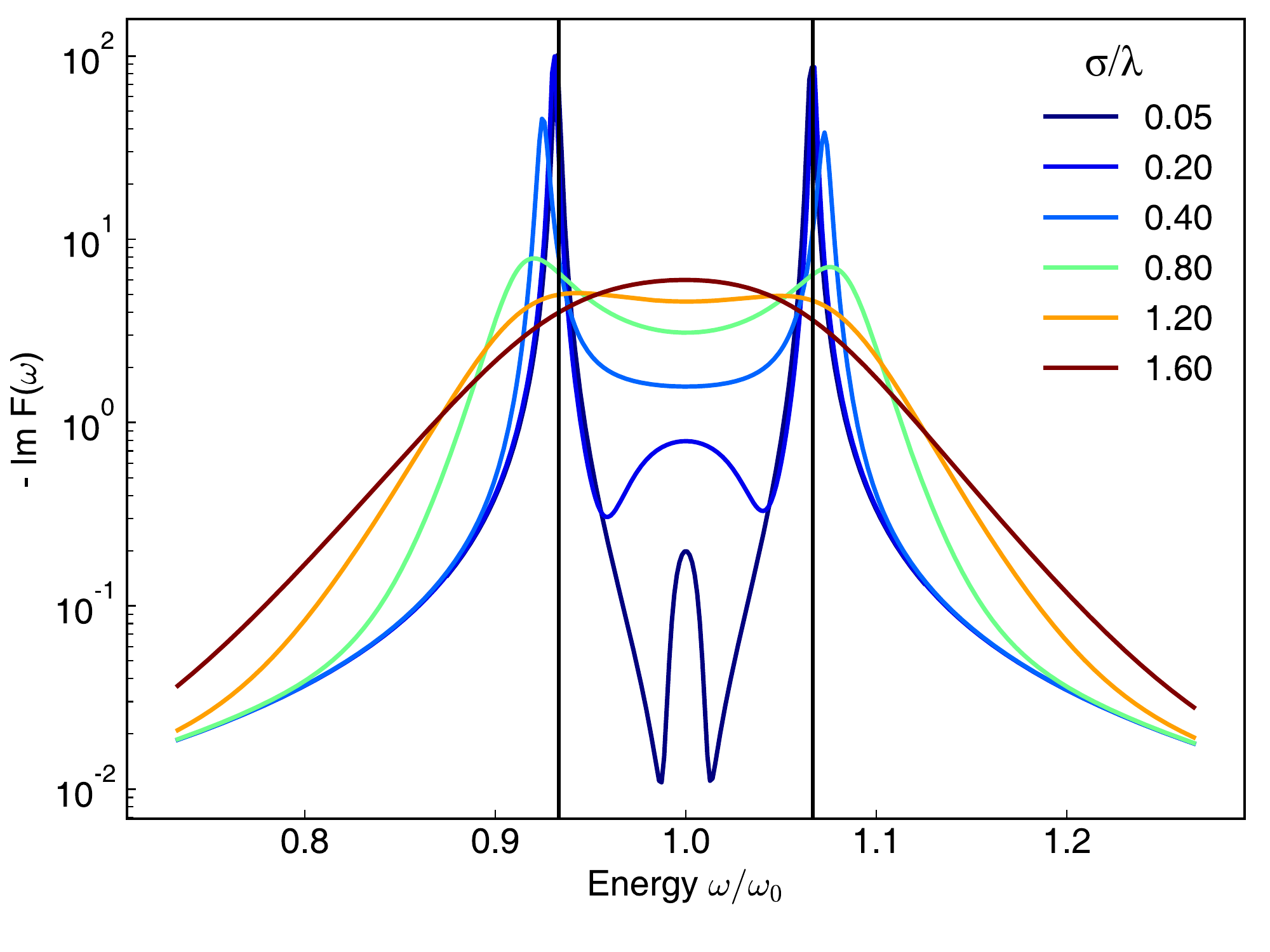}
	\caption{Photon Green's Function of an molecular ensemble consisting of $N$ two-level systems interacting with an cavity photon mode. The vertical lines indicate the energies of the disorder-free polaritonic states. As shown, the relationship between Rabi splitting and the static disorder is highly nonlinear. The vacuum Rabi splitting first increases while increasing the inhomogenous broadening in the transition energies of molecules when the $\sigma \ll g$, and then  decreases when $\sigma \sim \lambda$. The Rabi splitting eventually vanishes when $\sigma \gg \lambda$.}
	\label{fig:gf}
\end{figure}

\Fig{fig:gf} shows the photon GF in \eq{eq:pgf} at different disorder strength. As shown, the relationship between the splitting in the photon GF and the static disorder is highly nonlinear. The splitting first increases while increasing the inhomogenous broadening in the transition energies of molecules when the $\sigma \ll \lambda$, and then  decreases when $\sigma \sim \lambda$. The dark states acquire photonic fractions in this regime manifested as a peak at the center of the molecular transition frequency. The Rabi splitting eventually vanishes when $\sigma \gg \lambda$.

%

\section{Conclusions and Outlook} \label{sec:summary}
We have developed a general and exact pseduoparticle non-equilibrium Green's function theoretical framework,  to study the cooperative polaritonic dynamics in the strong coupling regime. This theory is valid for all molecular processes irrespective of the specific  Hamiltonian under study. In the thermodynamic limit,  this theory provides a mean-field-like picture whereby each molecules interacts with the local dressed cavity field. Finite $N$ corrections can be systematically included by a diagrammatic analysis.
%
%
We have implemented this theory and illustrated its utility for computing the photonic and molecular dynamics of the driven Tavis-Cummings model. It is further shown how to incorporate the photon leakage and energetic disorder into the dynamics.


This theory allows an exact modeling of the cooperative polaritonic quantum dynamics in the collective strong coupling regime and paves the way to understand cavity-altered molecular processes. The size of the quasiparticle GFs  depends on the number of vibronic states required to describe the dynamical process of interest. This will grow quickly with the molecular size. A realistic approach to treat large molecules is to use models with reduced dimensionality, which is widely used in studying photochemistry and spectroscopy \cite{hahn2000}. This direction is under progress.



%
%

\begin{acknowledgments}
	We thank Juan B. Pérez-Sánchez,  Dr. Yao Wang, and Yu Su for useful discussions.
\end{acknowledgments}

\bibliography{../../cavity,negf,optics,qchem}


\end{document}